\documentclass[english,aps,twocolumn,nofootinbib]{revtex4}
\usepackage[T1]{fontenc}
\usepackage[latin9]{inputenc}
\usepackage{wrapfig}
\usepackage{graphicx}
\usepackage{hyperref}
\usepackage{float}

\makeatletter
\@ifundefined{textcolor}{}
{%
 \definecolor{BLACK}{gray}{0}
 \definecolor{WHITE}{gray}{1}
 \definecolor{RED}{rgb}{1,0,0}
 \definecolor{GREEN}{rgb}{0,1,0}
 \definecolor{BLUE}{rgb}{0,0,1}
 \definecolor{CYAN}{cmyk}{1,0,0,0}
 \definecolor{MAGENTA}{cmyk}{0,1,0,0}
 \definecolor{YELLOW}{cmyk}{0,0,1,0}
}

\makeatother

\usepackage{babel}
\begin{document}

\title{Single-atom Transistors for classical computing}

\author{Huu Chuong Nguyen}
\email{huu.c.nguyen@student.uclouvain.be}
\affiliation{Université Catholique de Louvain}

\author{Maurice Retout}
\email{mretout@ulb.ac.be}
\affiliation{Université Libre de Bruxelles}

\author{Gildas Lepennetier}
\email{Gildas.Lepennetier@tum.de}
\affiliation{Technische Universität München}
\begin{abstract}
In this review, we present an overview of four proof-of-concept of single-atom transistors based on four technologies : Atom doping, Single electron transistors, Single-atom metallic wire and Multilevel atomic-scale switching. Techniques and methods to build these transistors are described as such as the theory behind their mechanisms.
\end{abstract}
\maketitle

\section{Introduction}
\begin{figure}
\includegraphics[width=0.75\columnwidth]{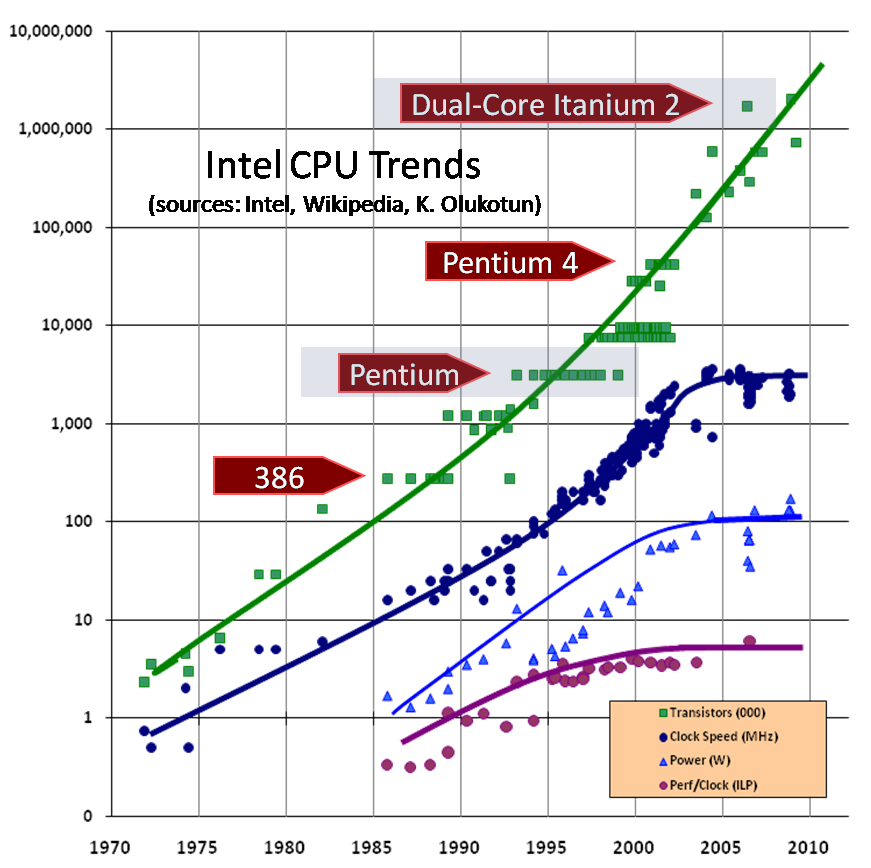}
\caption{CPU scaling showing transistor density, power consumption, and efficiency from \cite{Dobb}.}
 \label{CPU_fig}
\end{figure}

During the continuous improvement of computers' performance,
transistors have followed
the trend of Moore's law \cite{Moore} by doubling their number in integrated circuits approximately
every two years since 1965 (green curve on Figure \ref{CPU_fig}).  To increase their numbers on computer chips, transistors became smaller and smaller.
 Another trend linked to computer performance that followed the Moore law was the clock speed of the central processing unit (CPU), that indicates the number of operation performed by a chip per unit of time. 
 There are two main issues regarding these trends. By becoming increasingly smaller, transistors are more prone to quantum tunnelling. Additionnaly,  increasing the frequency of transistors also produce more heat\cite{Dobb}. Both effects lead to current leakage problems which is such an issue that manufacturer have to limit the clock speed of their transistors. It can be seen as a plateau of the dark blue curve on Figure \ref{CPU_fig}. 
The origin of these phenomena comes from the behaviour of the electron and its quantum properties. Since the shrinking of the transistors are bound to reach the atomic size at some points, quantum effects will be unavoidable and it is therefore important to understand them to design
new transistors. 
 
This review compiles the work of research groups that have developed so-called \emph{single-atom transistors} (SAT). 
We first present a short introduction on the basic of transistors in section \ref{basic}. 
We first briefly introduce the basic of transistors in section \ref{basic}.
Then, we present experimental techniques used to build SAT in section \ref{Techniques}. In section \ref{Principles}, we describe the principles behind the technologies of four families of SAT. And finally in section \ref{summary}, we conclude this review with a summary of the performances of these SAT and their limitations.
It is worth mentioning
that this review does not include SAT not yet
tested experimentally or solely designed  for quantum computing\cite{1D optical,quantumstorage,qbit,qbit2}.

\section{Basic of Transistors}
\label{basic}
\begin{figure}
\includegraphics[width=0.8\columnwidth]{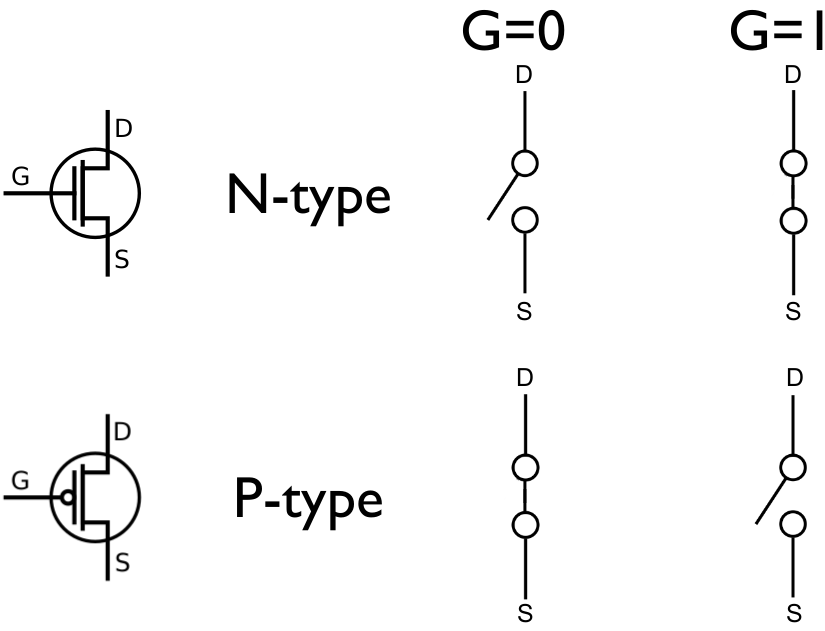}
\caption{MOSFET representations. N-type  (top) \& P-type  (bottom). At low voltage in the Gate N-type acts like an OFF switch and P-type as an ON Switch. At high voltage in the Gate, N-type acts like  a ON switch and P-type as an OFF switch.}
 \label{MOSFET_fig}
\end{figure}

Classical computers uses transistors to perform logical operations with 0 and 1. 
To be considered as a transistor, 
the device must act like a switch as shown in Figure \ref{MOSFET_fig}. Switching between an OFF and an ON state must be reproducible and controlled. 
The transistors presented in this review have three terminals, like
most modern transistors\cite{FET}:

\begin{itemize}  
\item Source (S): The terminal through which the electrons enter.
\item Drain (D): The terminal through which the electrons exit.
\item Gate (G): The terminal that modulates the channel's conductivity.
\end{itemize}

Semiconductors such as silicon have a conduction band, a valence band and a forbidden energy gap in between. Bands are in fact closely packed energy levels where electrons can stay. The valence band is usually filled so no electron in this band can move while the conduction band is at higher energy and is only partially filled.  In a solid, this allows the electrons from the conduction band to move and participate in a current flow. In reality, the valence is completely filled only at 0°K. Since energy level occupation follow the Fermi-Dirac distribution, when the temperature increases the valence band is less filled which can cause current leakage.

Transistors made of semiconductors materials usually need to be doped in order to obtain the desired electronic properties\cite{doping}. Doping is an intentional introduction of impurities into the semiconductor for the purpose of modulating its electrical properties. 
The addition of a dopant 
has two possible outcomes: 1) more electrons - \emph{i.e.} more negative charges - can be added to create n-type transistors, rising the Fermi level\footnote{In a band structure picture, the Fermi level can be considered to be a hypothetical energy level of an electron, such that at thermodynamic equilibrium this energy level would have a 50\% probability of being occupied at any given time. In practice, doping add new available states in the forbidden gap of the semi-conductor shifting the Fermi level of the material.}
or 2) the creation of 'holes' that add positive charges to create p-type transistors, decreasing the Fermi level
. 
At the nanoscale, the exact number and position of the dopant atoms will determine the type of transistors \cite{Shinada2005}. 


The presented single-atom transistors do not necessarily need to be doped nor have all the three terminals contained inside a single atom. They work thanks to very interesting phenomena appearing at the nanoscale. However, 
before talking about the mechanisms of these single-atom transistors, we have to investigate the methods and techniques used to build them.
%
Photolithography is a commonly used method to build commercial transistors but it does not suit the requirements for single-atom transistors because it does not allow to move accurately a single atom.
\cite{methods}.  Research groups have used different approaches described below to build their prototypes.  

\section{Building Techniques}
\label{Techniques}
Each research group used various techniques to build and determine
the properties of their prototypes. Here we present the main tools and principles common to most research groups. A detailed
review of these methods can be found in \cite{methods}.
\subsection{Electromigration}
Electromigration is the result of momentum transfer from the electrons, which move in the applied electric field, to the ions contained in the lattice of the material. 
Uncontrolled electromigration can cause several kinds of failures in electronic component \cite{EM1,EM2}. However, by carefully tuning the current, electromigation can be used to move atoms towards the desired place \cite{EM3}.

\subsection{Scanning Tunneling Microscope}
The Scanning Tunneling Microscope (STM) allows to measure the conductance of a molecule and to \emph{see} the atoms at a resolution of 0.1 nm. 
The image of the probed surface is reconstructed by analysing the current flowing through the STM tip \cite{STM,STM2} which is sometime called STM topography. They can also be used on quatum points contact as a two-terminal conductance switching devices \cite{STMswitch} and move atoms via electromigration which is sometime called STM lithography \cite{Psingle1,Psingle}.

\subsection{Molecular Break-junction}
\begin{figure}[h]
\begin{center}
      \includegraphics[angle=0,width=0.61\columnwidth]{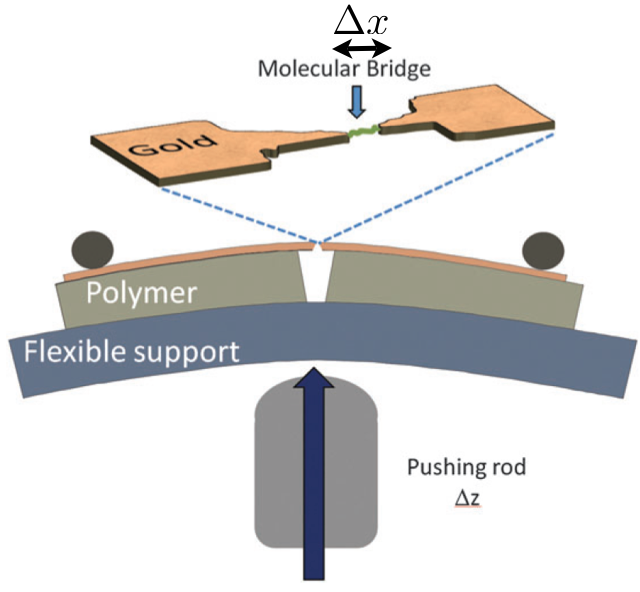}
  \caption{Schematic of a molecular break-junction.}
  \label{junction_fig}
\end{center}
\end{figure}

The Molecular break-junction is a technique where a thin metallic wire on a flexible support is bent until it breaks  \cite{Roch2008}. The separated wires form the electrodes and the gap between them is then bridged by a covalently bonded 
molecule to form a molecular junction. By controlling the pushing ($\Delta z$), the gap width
($\Delta x$) can be very well controlled with a precision up to a picometre ($10^{-12}$~m) \cite{PhysRevLett.99.026601} (See Figure \ref{junction_fig}).
The part of the wire that will break is usually immersed in a solution that contains the molecules of interest that will anchor to the wire once it is broken.

\subsection{Electrochemical deposition}
The Electrochemical deposition technique uses a device similar to molecular break-junction after breaking the metallic wire but without mechanical pushing rod. The two electrodes are separated by a gap and are in a solution containing the ions of interest. By controlling the potential between the electrodes, the ions will aggregate on the electrode surface. 
The  width  of  the  constriction  can  be  adjusted  by
slowly  dissolving  metal  atoms  away  or  redepositing  atoms  onto  the  constriction  which  can  be controlled solely by the electrode potential. This method avoids the mechanical stress induced in Molecular Break-junction that can cause unwanted defects.



\section{Single-Atom Transistors Principles}
\label{Principles}
Controlling the charge transfer on a single atom can be very difficult
\cite{AuCharge}. Here we present 4 types of transistors based on
different technologies.

\subsection{Single-Atom doping}
\begin{figure}[h]
\begin{center}
      \includegraphics[angle=0,width=0.55\columnwidth]{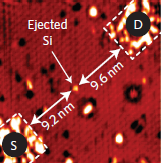}
  \caption{STM picture of the system doped with a single phosphorus atom from \cite{Psingle}. The ejected Si atom is replaced by a single P atom in the lattice. }
  \label{doping_fig1}
\end{center}
\end{figure}

Several research groups were able to make transistors by adding a single doping atom. Two research groups \cite{Psingle1,Psingle} were able to accurately place a single doping phosporus  atom on H-passivated Si surface using STM lithography. This resulted in a n-type doping since phosphorus has more valence electrons than silicium. 
The method used is a four-steps process : (i) The STM removes H atoms around the site where we want the dopant to be, (ii) PH3 gas is injected at room temperature and is fixed to the surface as PH2, (iii) the surface is heated at 350 °C until a single P atom is incorporated to the surface layer which cause the ejection of a Si atom and remove the excess of P atoms (See Figure \ref{doping_fig1}).
Other research groups \cite{Haider2009,Bellec2010,SiSingle} have used other elements such as Gallium (Ga) to study how this p-type doping affect the behaviour of a single Si dangling bond using similar techniques to build their systems.

\begin{figure}[h]
\begin{center}
      \includegraphics[angle=0,width=0.95\columnwidth]{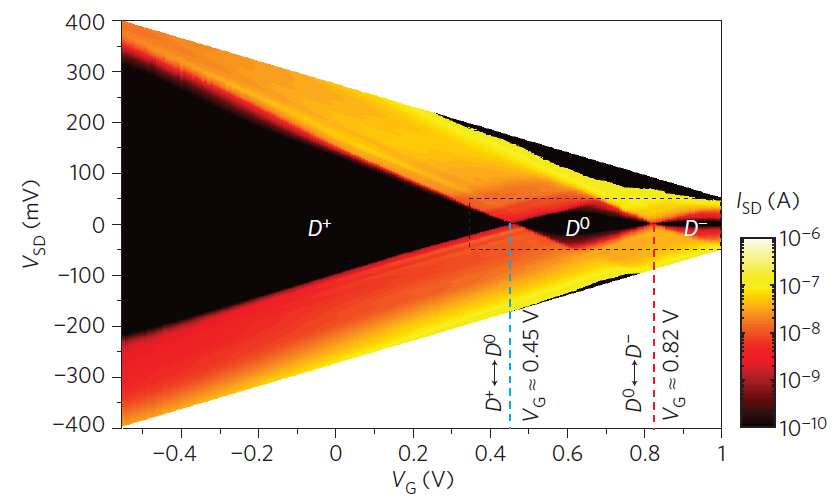}
  \caption{Typical stability diagram of a single-atom doped transistor from \cite{Psingle} showing the current in function of the gate voltage and source-drain voltage. The system here correspond to a single phosphorus atom doped on epitaxial Si surface. $(D^+,D^0,D^-)$ correspond to the (0,1,2) electron states in the phosphorus atom.}
  \label{doping_fig2}
\end{center}
\end{figure}

Each research group had to ensure that their devices operate correctly and was 
stable under working conditions. They used either Tight-binding\footnote{Tight-Binding \cite{TB} is a semi-empirical model where the full Hamiltonian of the system is approximated by the Hamiltonian for isolated atoms 
by using a Hartree-product ansatz. Then we assume that the eigenfunctions of the single-atom Hamiltonian are very small at distances exceeding the lattice constant, so that the lattice sites can be treated independently. One important point in the tight-binding model is that the electrons behave in a solid almost as if they were in vacuum but with a different mass. This leads to the \emph{effective mass equation} which is very similar to the Schrödinger equation}
or Density Functional Theory\footnote{The Kohn-Sham formulation of Density Functional Theory (DFT) \cite{PhysRev.140.A1133} allows to map a full interacting system with the real $N$-body potential from the Schrödinger equation onto a fictitious non-interacting system whereby the electrons move within an effective Kohn-Sham single-particle potential. The Kohn-Sham method is still exact since it yields the same ground state density as the real system.} simulations
  to predict the behaviour of their devices and then experimental measurements 
to check that their devices operated as designed. These devices are usually studied at a temperature close to 0°K. This has to ensure that the extra discrete energy level coming from the doping will be visible and that the valence band is always full (energy levels' occupancy follows the Fermi-Dirac distribution) to avoid unwanted tunnelling current. The experiments done by each groups were in good agreement with their simulations as it showed new available discrete energy level as one can expect from single-atom doping (See Figure\ref{doping_fig2}). 

\subsection{Single electron transistor}

\begin{figure}[h]
\begin{center}
            \includegraphics[angle=0,width=0.45\columnwidth]{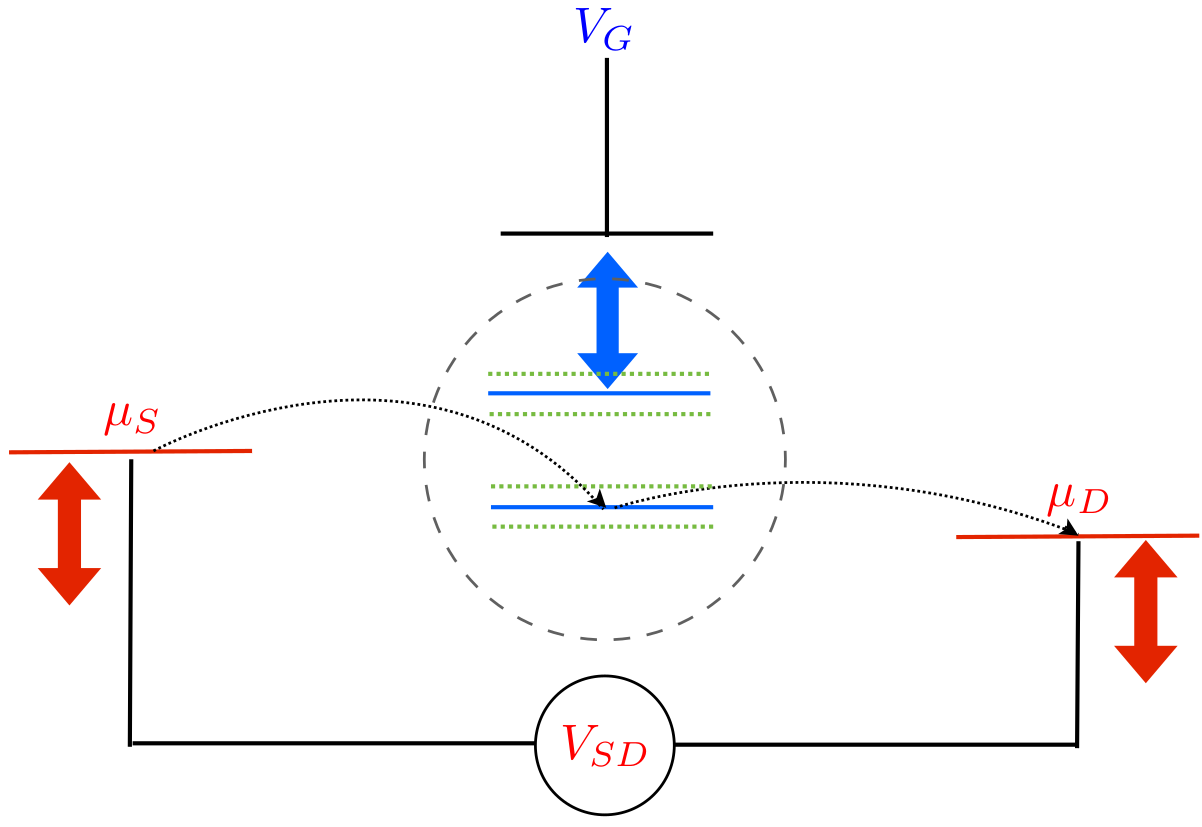} \, \includegraphics[angle=0,width=0.45\columnwidth]{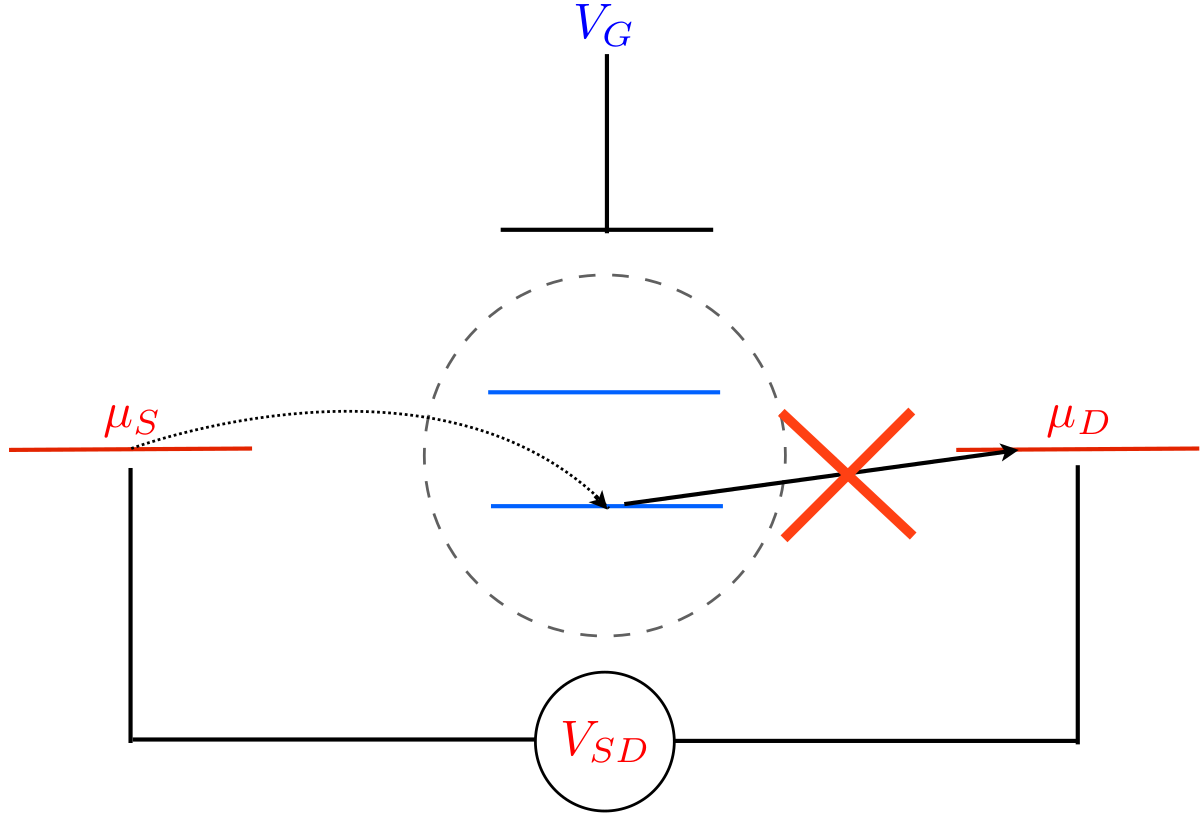}    
  \caption{Schematical view of Coulomb blockade principle in a SET. The voltage of the electrode $V_{SD}$ increase or decrease the chemical potentials of the source $\mu_S$ and drain $\mu_D$ (red). The gate voltage influences the energy levels inside the QD (blue). The Zeeman effect can split energy levels in the QD (green). Current can pass if and only if there is at least an energy level between $\mu_S$ and $\mu_D$.}
  \label{SET}
\end{center}
\end{figure}

Single-electron transistor (SET) \cite{SingleElec1,SingleElec2} are devices that have a source, a drain and in the center, a quantum dot (QD) sometimes called artificial atom because it has clearly separated energy levels. 
Working at temperature close to 0K, for the same reasons as in single-atom doped transistors,  
the current will flow only if the electrons from the source have an energy above the drain and also at least one available energy level in the QD between the energy levels of the electrodes. When there are no energy levels available, 
the QD acts as a off-switch and no current can pass through. This phenomenon is called Coulomb Blockade. More explicitly, when an electron leave the source, it has an energy equals to $\mu_S$, it can only occupy energy levels of the QD, $E_n$ smaller or equal to $\mu_S$. Once in the dot, the electron will reach the drain only if the chemical potential of the drain $\mu_D$ is smaller than the occupied energy in the dot. This can be summarize with the relation $\mu_S>E_n>\mu_D$ (see Figure \ref{SET}).

To turn on or off the QD, one can tune the source-drain potential, $V_{SD}$ that change the chemical potential ($\mu_S$ and $\mu_D$) (red in Figure \ref{SET} left) of the leads and the potential of the gate, $V_G$ (blue in Figure \ref{SET} left) that influences the energy levels of the QD. 

\begin{figure}[h]
\begin{center}
      \includegraphics[angle=0,width=0.95\columnwidth]{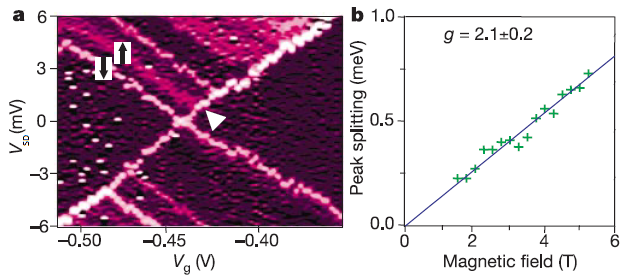}
  \caption{Edited picture from \cite{Coulomb} showing (a) Stability diagram of their SET under a magnetic field of 6T. New energy level due to Zeeman effect is indicated with the triangle. (b) Magnitude the Zeeman spliting in function of the magnetic field}
  \label{Zeeman}
\end{center}
\end{figure}

Another source of energy levels come from the spin interaction with an external magnetic field. Electrons with spin up or down behaves differently under the same magnetic field. In practice, it is the Zeeman effect that lift the degeneracy of the energy levels \cite{Coulomb}. This creates additional energy levels available in the QD (Green doted lines in Figure \ref{SET}), that can be used for a transistor as it had been observed experimentally in \cite{Coulomb} (see Figure \ref{Zeeman}). A correct theoretical treatment of this effect requires to change the Schrödinger equation to include spin-orbit coupling or the use the Dirac equation that naturally includes them \cite{Dirac}.

\begin{figure}[h]
\begin{center}
      \includegraphics[angle=0,width=0.94\columnwidth]{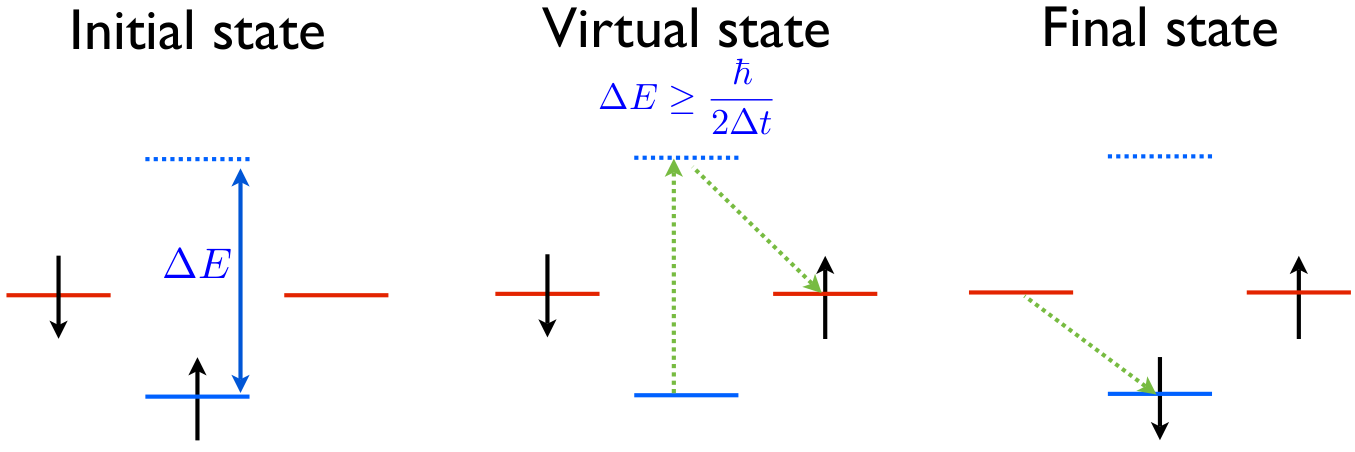}
  \caption{Schematical representation of the Kondo resonance. Initial state is a Coulomb blockade situation. During the virtual state, the electron in the QD can transit in an excited state during a very short time $\Delta t$ and exit to the drain. The empty level in the QD is filled by the electron from the source.}
  \label{Kondo_resonance_fig}
\end{center}
\end{figure}

\begin{figure}[h]
\begin{center}
      \includegraphics[angle=0,width=0.95\columnwidth]{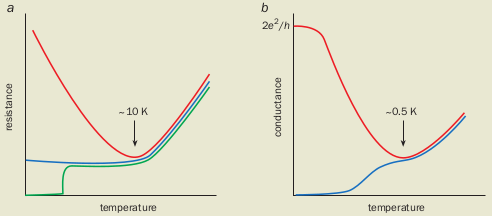}
  \caption{(a) As the temperature of a metal is lowered, its resistance decreases until it saturates at some residual value (blue). Some metals become superconducting at a critical temperature (green). For metals that contain magnetic impurities, the resistance increases at low temperatures due to the Kondo effect (red). (b) Conductance of a QD in function of the temperature. The Kondo effect is present when the number of electron is odd in a QD (red). The Kondo effect does not occur when the dot contains an even number of electrons in the QD (blue). Picture edited from \cite{Kondo}.}
  \label{Kondo_Metal_QD}
\end{center}
\end{figure}

Finally, there is the effect of magnetic impurities at low temperature called Kondo effect.
At very low temperature (below 10 K) some metal like Pb, Nb, Al for example lose all resistance and become superconducting. While others such as Cu or Au have a finite resistance and we observe that their resistivity increases with the number of magnetic defects present (see Figure \ref{Kondo_Metal_QD}a).
The origin of this phenomenon is the Kondo resonance\footnote{The description of the Kondo resonance relies on the perturbation theory\cite{Kondo_reso} that describes the electron conduction as we approach the temperature of 0 K.} which is the coupling of an unpaired electron in the impurity and an electron in the surrounding.
The Kondo resonance allows electron transfer from the QD when it would be normally forbidden in a Coulomb blockade situation. The mechanism can be described in its simplest case as follow (see Figure \ref{Kondo_resonance_fig}). 

(i) For a given electron with a given spin in the QD coupled to the electron in the source that has the opposite spin as initial state. It is in a Coulomb blockade situation since the energy level of the electron in the QD is lower than the energy level of the source and drain. (ii) The Heisenberg incertitude principle applies on the energy of the energy level of the electron in the dot. It is possible for the electron in the dot to reach an excited energy level that is above the level of the drain and therefore exit in the drain. This virtual state is only possible for a very short amount of time with respect to the Heisenberg incertitude principle $\Delta E \cdot \Delta t \geq \hbar/2$. (iii) The dot has an available energy level that will be filled by the paired electron from the drain in the final state. This results in the spin switch from the electron inside the QD and the electron in the leads.

One of the main distinctions between a quantum dot and a macroscopic bulk metal is  their different geometries. In a metal, the electron states are plane waves, and scattering from impurities in the metal mixes electron waves with different momenta in any direction. This momentum transfer increases the resistance.
In a quantum dot, however, all the electrons have to travel through the device, as there is no other electrical path around it. In this case, the 
Kondo resonance 
makes it easier for states belonging to the two opposite electrodes to mix. This mixing increases the conductance (\emph{i.e.} decreases the resistance). In other words, the Kondo effect produces the opposite behaviour in a quantum dot to the one in a bulk metal \cite{Coulomb,Kondo,Kondo1,Kondo2,KondoMol}.
If the number of electron in the QD ($N$) is odd there is an unpaired electron with a free spin
which can form a singlet with electrons at the Fermi level in the leads. This coupling results in a greater density of states at the Fermi level of the leads \cite{PhyRevB14782}. 
The Kondo effect does not occur when the dot contains an even number of electrons and the total spin adds up to zero \cite{Kondo} (see Figure \ref{Kondo_Metal_QD}b). Increasing the temperature destroys the singlet and attenuates the conductance.

Such single electron transistors have effectively been built using Molecular Break-junction and STM lithography techniques with a single atom of Cobalt \cite{Coulomb} or Arsenic \cite{Kondo2} as the QD. The coulomb blockade and Kondo effect have been observed using these single gate-tunable atom.


\subsection{Single-Atom metallic wire}
The transistors presented above combine electronic components made of different kinds of elements with different electronic properties so that they \emph{act} as switch when put together. As each part was connected, low temperature was necessary to avoid unwanted tunnelling current.

\begin{figure}[h]
\begin{center}
      \includegraphics[angle=0,width=0.9\columnwidth]{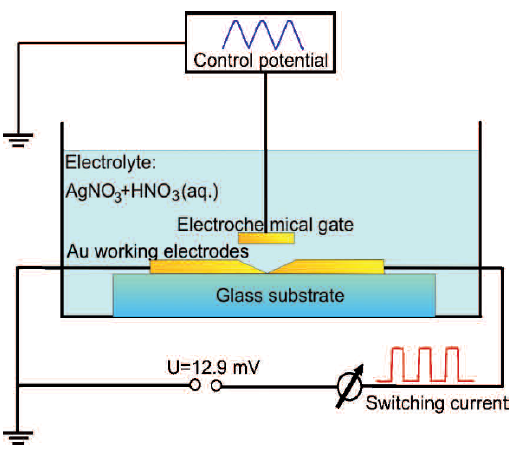}
  \caption{Schematical representation of the Single-Atom metallic wire device used in \cite{Wire1,wire2,wire3,wire4,wire5,wire5a,wire6}. Edited from \cite{wire2}}
  \label{Ag_wire_device}
\end{center}
\end{figure}
Some groups \cite{Wire1,wire2,wire3,wire4,wire5,wire5a,wire6,wirePb} have taken a completely different approach as they have successfully managed to create single-atom transistors made of pure metals that behave even more like an everyday common switch since they effectively close/open the wire to switch on/off the current. 

\begin{figure}[h]
\begin{center}
      \includegraphics[angle=0,width=0.9\columnwidth]{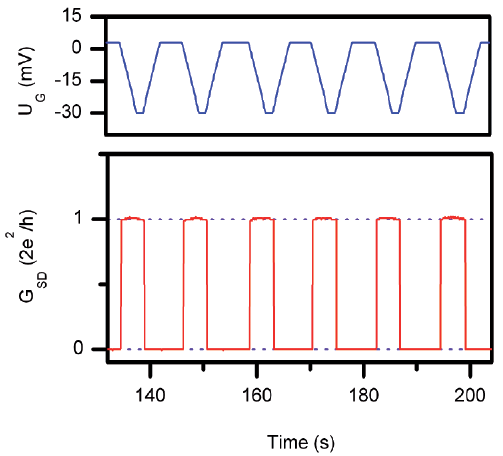}
  \caption{Experimental measurement of the source-drain
conductance ($G_{SD}$)(red
curves) is directly
controlled by the
gate potential ($U_G$)
(blue curves) in \cite{Wire1,wire2,wire3,wire4,wire5,wire5a,wire6}.}
  \label{UgvsG}
\end{center}
\end{figure}

To build their devices, they have used electrochemical deposition techniques \cite{Tao1998,Elhoussine2002} (see Figure \ref{Ag_wire_device}). Starting from a gold electrodes spaced by about 100 nm in a solution containing either silver (Ag) \cite{Wire1,wire2,wire3,wire4,wire5,wire5a,wire6} or lead (Pb) \cite{wirePb}, the metallic atom will fill the gap gradually.
The newly formed wire is made to have only a single bridging atom as contact. This can be ensured experimentally as the conductance of a metallic wire made of a single atom is equal to the quantum of conductance \cite{Xie2008},  $G_0=2e^2/h$ where $e$ is the electron charge and $h$ the Planck's constant. Finally, a gate is used to control the single bridging atom. Changing the potential of the gate $U_g$  changes the measured Source-Drain conductance from 0 to 1 $G_0$ ($G_{SD}$) as shown in Figure \ref{UgvsG}. Their results indicate that the gate effectively move the bridging atom and can reproducibly operates at room temperature.



\subsection{Multilevel atomic-scale switching}
\begin{figure}[h]
\begin{center}
      \includegraphics[angle=0,width=0.95\columnwidth]{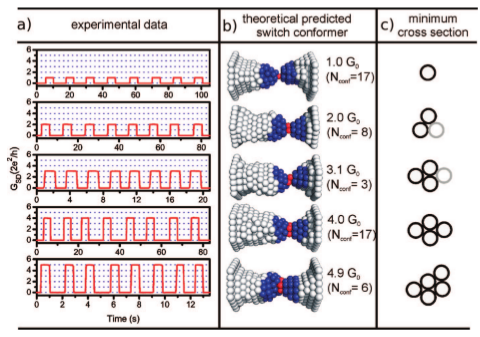}
  \caption{Picture from \cite{wire4} showing the relationship between the structures of atomic point contacts and their conductance. (a) is the measured source drain conductance. (b) Representative conformations of simulated junctions and number of junctions with the specified conductance. (c) Representative minimal cross-sections for each conductance level.}
  \label{Agconfig}
\end{center}
\end{figure}

Using the same Single-Atom metallic wire devices, described in the previous section, made of Ag \cite{wire2,multi1,multi1a,multi2,multi3} or Al \cite{multi3}, research groups have shown that it is possible to switch between more than only two on/off states. Simulations and measurements demonstrated that there are multiple stable geometrical conformations as shown in Figure \ref{Agconfig}b that have different quantized conducting states (see Figure \ref{Agconfig}a) depending on the number of atoms that make the cross section of the wire (see Figure \ref{Agconfig}c).  

\begin{figure}[H]
\begin{center}
      \includegraphics[angle=0,width=0.9\columnwidth]{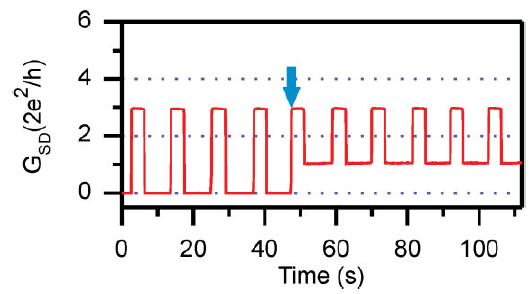}
  \caption{Experimental demonstration from \cite{multi1a} of multilevel atomic transistors switching an \emph{off state} and two \emph{on states}.}
  \label{multi}
\end{center}
\end{figure}

It is possible to switch from an \emph{off state} were the conductance is equal to 0 to any of these conducting states but also from one of these states to another as shown on Figure \ref{multi}. The last possibility does not require the contact to be broken and operates as a \emph{multi-level atomic quantum transistor}. Such multi-level logics and storage devices on the atomic scale would be of great interest as they could allow a more efficient data storage and processing with a smaller number of logical gates. They are however much slower than conventional transistors since their frequency is much lower. 

\section{Limitations and conclusions}
\label{summary}
We have presented four proof-of-concept technologies of transistors at the atomic scale. SAT based on \emph{Single-Atom doping} and \emph{Single electron transistor} can operate at maximum frequency but requires a constant cooling as they can operate only at very low temperature.
SAT based on {Single-Atom metallic wire} and \emph{Multilevel atomic-scale switching} operate at room temperature but require to be in an ionic solution and operate only at very low frequencies that might be compensated by multi-level logic gates.
 However, they are still far from immediate commercial use since experimental conditions differ greatly from every day normal conditions (temperature, pressure, environment,...) and the techniques used to build them is currently too expensive for mass production compared to today commercial semi-conductor transistors
\cite{limits}. 

Despite these facts, each presented SAT represent a technological breakthrough since it operates as a switch and is stable under working conditions as intended.

\begin{acknowledgments}
This review was written under the supervision of Prof. Bayot Vincent
in the context of the course LELEC2550 - \textit{Special semiconductor
devices}.
\end{acknowledgments}


\begin{thebibliography}{10}
\bibitem{Moore}Gordon E. Moore. (1965), 
Electronics, pp. 114\textendash 117.

\bibitem{Dobb} Dr. Dobb's Journal, 30(3), March 2005.

\bibitem{FET}Jacob Millman \& Christos C. Halkias .(1968),\textit{Electronic devices and circuits}. 
McGraw-Hill. pp. 384\textendash 385,
. ISBN 0-07-085505-6.

\bibitem{methods}Agraït, N., Yeyati, A. L., \& VanRuitenbeek, J.
M. (2003). 
Physics Reports, 377(2\textendash 3), 81\textendash 279. 

\bibitem{STM}Binnig, G.; Rohrer, H. (1986), 
IBM Journal of Research and Development. 30 (4): 355-69.

\bibitem{STM2}G. Schitter; M. J. Rost (2008), 
Mater. Today, 11 Supplement(0):40-48 

\bibitem{STMswitch} D.P.E. Smith. (1995), 
Science 269, 371.

\bibitem{EM1} Ogurtani, T. O. (2009), 
J. Appl. Phys., 106(5).

\bibitem{EM2} J. R. Lloyd. (1991), 
J. Appl. Phys. 69, 7601 

\bibitem{EM3}Terabe, K. et al. (2005). 
Nature, 433(7021), 47.  

\bibitem{Roch2008} Roch, N., et al. (2008), 
Nature, 453(7195), 633.

\bibitem{PhysRevLett.99.026601} Parks, J. J. et al. (2007), 
 Phys. Rev. Lett. 99, 026601.
33.
\bibitem{1D optical}Micheli, A., Daley, A. J., Jaksch, D., \& Zoller,
P. (2004), 
Phys. Rev. Lett., 93(14), 2\textendash 5

\bibitem{quantumstorage}Loth, S., Etzkorn, M., Lutz, C. P., Eigler,
D. M., \& Heinrich, A. J. (2010), 
Science, 329(5999),
1628\textendash 1630.

\bibitem{qbit}Prati, E., Belli, M., Cocco, S., Petretto, G., \& Fanciulli,
M. (2011), 
Appl. Phys. Lett., 98(5), 2009\textendash 2012.

\bibitem{qbit2}Vincent, R., Klyatskaya, S., Ruben, M., Wernsdorfer,
W., \& Balestro, F. (2012), 
Nature, 488(7411), 357\textendash 360. 

\bibitem{AuCharge}Repp, J. (2004), 
Science, 305(5683), 493\textendash 495. 

\bibitem{doping}Deen, William M. (1998). \textit{Analysis of Transport Phenomena}. pp. 91-94.

\bibitem{Shinada2005}Shinada, T. et al. (2005), 
Nature, 437, 1128-1131. 

\bibitem{PhysRev.140.A1133}Kohn, W., \& Sham, L. J. (1965), 
Phys. Rev., 140(4A), A1133.

\bibitem{SiSingle}Bellec, A., Chaput, L., Dujardin, G., Riedel, D.,
Stauffer, L., \& Sonnet, P. (2013),
Phys. Rev. B, 88(24), 241406. 

\bibitem{TB}  Slater, J. C. \&  Koster,G. F. (1954), 
 Phys. Rev. 94, 1498

\bibitem{Psingle}Fuechsle, M. et al.. (2012), 
Nature Nanotechnology, 7(4), 242\textendash 246. 

\bibitem{Psingle1} Schofield, S. R. et al. (2003), 
Phys. Rev. Lett. 91, 136104.

\bibitem{Haider2009}Haider et al. (2009), 
 Phys. Rev. Lett., 102(4), 46805.

\bibitem{Bellec2010}Bellec, A. et al. (2010), 
Phys. Rev. Lett., 105(4), 48302. 

\bibitem{SingleElec1}Baldea. (1998), 
Nature, 391(January), 156\textendash 159.

\bibitem{SingleElec2}Kastner, M. A. (2000), 
Annalen Der Physik (Leipzig), 9(11\textendash 12),
885\textendash 894. 

\bibitem{Coulomb}Park, J., et al. (2002), 
Nature, 417(6890),
722\textendash 725.

\bibitem{Dirac}Dirac, P. A. M. (1928), 
Proc. R. Soc. A. 117 (778), 610.

\bibitem{Kondo_reso} Kondo, J. (1964) Prog. Theor. Phys., 32 (1), 37-49. 

\bibitem{Kondo1}van der Wiel, W. G. (2000), 
Science, 289(5487), 2105\textendash 2108.

\bibitem{Kondo}Kouwenhoven, L., \& Glazman, L. (2001), 
Phys. World, 14(1), 33\textendash 38. 

\bibitem{KondoMol}Liang, W., Shores, M. P., Bockrath, M., Long, J.
R., \& Park, H. (2002), 
Nature, 417(6890), 725\textendash 729.

\bibitem{Kondo2}Lansbergen, G. et al. (2010), 
Nano Lett., 10(2), 455\textendash 460. 

\bibitem{PhyRevB14782}Wan, Y. et al. (1995), 
Phys. Rev. B, 51(20), 14782-14785.


\bibitem{Wire1}Maul, R., Xie, F.-Q., Obermair, C., Schön, G., Schimmel,
T., \& Wenzel, W. (2012), 
Appl. Phys. Lett., 100(20),
203511. 

\bibitem{wire2}Obermair, C., Xie, F.-Q., \& Schimmel, T. (2010),
Europhysics News, 41, 25\textendash 28. 

\bibitem{wire3}Obermair, C., Kuhn, H., \& Schimmel, T. (2011),  
Beilstein J. Nanotechnol.,
2(1), 740\textendash 745. 

\bibitem{wire4}Xie, F. Q., Maul et al. (2008), 
Nano Lett., 8(12), 4493\textendash 4497. 

\bibitem{wire5}Xie, F. Q., Nittler, L., Obermair, C., \& Schimmel,
T. (2004), 
Phys. Rev. Lett., 93(12), 1\textendash 4.

\bibitem{wire5a}Xie, F. Q., Obermair, C., \& Schimmel, T. (2004), %
Solid State Commun., 132(7),
437\textendash 442. 

\bibitem{wire6}Xie, F.-Q. et al. (2008), 
Appl. Phys. Lett., 93(4), 43103. 

\bibitem{wirePb}Xie, F.-Q., Hüser, F., Pauly, F., Obermair, C., Schön,
G., \& Schimmel, T. (2010), 
Phys. Rev. B, 82(7),
75417. 

\bibitem{Tao1998} Li, C. Z., \& Tao, N. J. (1998), 
Appl. Phys. Lett. 72, 894.

\bibitem{Elhoussine2002}Elhoussine, F. et al. (2002),  
Appl. Phys. Lett., 81, 1681.

\bibitem{Xie2008}Xie, F.-Q. et al. (2008), 
Appl. Phys. Lett., 93, 43103

\bibitem{multi1}Gleiter, H., Schimmel, T., \& Hahn, H. (2014),  
Nano Today, 9(1),
17\textendash 68. 

\bibitem{multi1a}Xie, F., Maul, R., Obermair, C., Wenzel, W., Schön,
G., \& Schimmel, T. (2010), 
?Adv. Mater., 22(18),
2033\textendash 2036. 

\bibitem{multi2}Olsson, F. E., Paavilainen, S., Persson, M., Repp,
J., \& Meyer, G. (2007), 
Phys. Rev. Lett., 98(17), 176803. 

\bibitem{multi3}Schirm, C., Matt, M., Pauly, F., Cuevas, J. C., Nielaba,
P., \& Scheer, E. (2013), 
 Nature Nanotech., 8(9), 645\textendash 648. 

\bibitem{limits}De Franceschi, S., \& Kouwenhoven, L. (2002), 
Nature, 417(6890), 701\textendash 702. 
\end{thebibliography}
\end{document}